\documentclass[AMA,STIX1COL]{WileyNJD-v2}
\usepackage{moreverb}
\usepackage{gensymb}
\usepackage{booktabs}
\usepackage{threeparttable}
\usepackage{amsmath}
\usepackage{array}
\usepackage{subfigure}
\usepackage{bm}
\usepackage{lineno}
\usepackage{cleveref}
\allowdisplaybreaks[4]

\newcommand\BibTeX{{\rmfamily B\kern-.05em \textsc{i\kern-.025em b}\kern-.08em
		T\kern-.1667em\lower.7ex\hbox{E}\kern-.125emX}}

\articletype{Original Paper}%

\received{<day> <Month>, <year>}
\revised{<day> <Month>, <year>}
\accepted{<day> <Month>, <year>}

\begin{document}
	\title{{Links Assignment Scheme based on Potential Edges Importance in Dual-layer Wavelength Routing Optical Satellite Networks}}
	
	\author[1]{Jingkai Yang*}
	\author[1]{Qiwen Ran}
	\author[1]{Hongyu Wu}
	\author[1]{Jing Ma}
	\authormark{Yang \textsc{et al}}
	
	\address[1]{\orgdiv{National Key Laboratory of Science and Technology on Tunable Laser},	\orgname{Harbin Institute of Technology}, \orgaddress{\state{Harbin}, \country{China}}}
	
	\corres{Jingkai Yang, National Key Laboratory of Science and Technology on Tunable Laser, Harbin Institute of Technology, Harbin, 150001, China.
		\email{ynyxyjk@stu.hit.edu.cn}}
	
	\abstract[Abstract]{With the development of the massive satellite constellation and the on-orbit laser-based communication equipment, the wavelength routing optical satellite network (WROSN) becomes a potential solution for on-orbit, high-capacity, and high-speed communication. Since the inter-satellite links (ISLs) are time-varying, one of the fundamental considerations in the construction of the WROSN is assigning limited laser communication terminals for each satellite to establish ISLs with the visible satellites. Therefore, we propose a links assignment scheme (LAS) based on the potential edges importance matrix (PEIM) algorithm to construct a temporarily stable topology of the ISLs for a dual-layer constellation. The simulation results showed that the LAS based on the PEIM algorithm is better than LAS based on the random or Greedy algorithm in terms of node-to-node distance, node pair connectivity, wavelength demand, and transmission delay. The node pair connectivity and wavelength demand in WROSN is a trade-off problem. The research in this paper also brings a novel method for reduction of the cost of the on-board resources, that is through designing topology of the ISLs with links assignment algorithm.}
	
	\jnlcitation{\cname{%
			\author{Yang J.}, 
			\author{Ran Q.}, 
			\author{Wu H.}, and 
			\author{Ma J.}} (\cyear{2023}), 
		\ctitle{Links Assignment Scheme based on Potential Edges Importance in Dual-layer Wavelength Routing Optical Satellite Networks}, \cjournal{Int J Satell Commun Network}, \cvol{2023}.}
	
	\maketitle
	
	
	\section{Introduction}
	Satellite networks have drawn much attention for their unique advantages in many applications, such as global communication, remote sensing, military activities, and so on. However, the growing demand for bandwidth and capacity for data and multimedia services has led to congestion in the conventionally used radio frequency spectrum and arises a need for high-capacity, high-speed satellite transport networks\cite{4063386,chan2003optical,7553489}. The key solution to meet this demand is to form an optical network over the satellite constellation\cite{https://doi.org/10.1002/sat.1351,9219130,9141375}. The constellation will consist of hundreds of appropriately designed satellites to provide seamless global coverage. In this system, several laser communication terminals (LCTs) will be deployed on each satellite to establish inter-satellite links (ISLs) with other satellites and support multi-hop relay services. Numerous national departments and commercial enterprises have already announced their plans for optical satellite network (OSN), such as NeLS\cite{suzuki2004study}, EDRS\cite{calzolaio2020edrs}, Starlink\cite{chaudhry2021laser} and Kuiper\cite{bhattacherjee2019network}.
	
	ISLs are hardly affected by atmospheric absorption, scatting, or turbulence, so the communication rate can reach several Gb/s. In addition, with the maturation of on-board wavelength division multiplex (WDM) technology and optical cross-connect devices, using WDM ISLs and wavelength routing technology to serve the communications between satellites is foreseeable. The wavelength routing OSN will have lower transmission latency, shorter processing time, larger capacity, and simpler routing decisions\cite{908734}. These advantages make it possible to build all-optical satellite backbone networks and provide high-quality services for the future Internet of Things and Space-Ground Integrated Networks (SGINs).
	
	One of the fundamental considerations in constructing the wavelength routing OSN is the design of the physical topology of the ISLs. The wavelength routing OSN over a single-layer low-earth orbit (LEO) satellite constellation was firstly proposed and analyzed in 2000\cite{908734}, and the goal is to construct a stable topology of the ISLs. The satellites use four LCTs to establish ISLs and communicate with two adjacent intraorbital satellites and with two adjacent orbits on both sides. This kind of regular bi-directional Manhattan Street Networks (MSNs) can be maintained permanently, and the performance of the OSN based on this design is analyzed in Next-Generation Low Earth Orbit System (NeLS) and Iridium constellations\cite{suzuki2004study,bhattacherjee2019network,5669633}. However, though the LEO satellites alleviate the delay problem, the massive multi-hop services may bring large queuing delay\cite{tan2014analysis}. Since the most of traffic in the OSN will be data-based, the effects on user quality of service of the geosynchronous (GEO) delay can be minimized\cite{chan2003optical}. Deploying OSN on the dual-layer constellation, which consists of several LEO and GEO satellites, has more advantages. The dual-layer constellation has better coverage for terrain users, and the GEO satellites can be the stable interface for terrain gateway stations.
	
	However, the links assignment problem in the dual-layer wavelength routing OSN (DWROSN) is more complicated than in the single-layer constellation. On the one hand, the high orbital altitude satellites have numerous line-of-sight low orbital altitude satellites, which brings a variety of potential link selections. On the other hand, the impermanent ISLs make the link topology changes dynamically and frequently, which affects the topology of the ISLs and the routing and wavelength assignment (RWA) directly. These features deteriorate the design of the links assignment scheme (LAS) in DWROSN. Focus on topology design problem of the ISLs in the single-layer OSN, researches have been proposed\cite{5669633,10.1007/978-3-030-19153-5_61,liu2020analytic}. The research of the randomly connected ISLs in the OSN analyzes the effect of terminal utilization on wavelength requirement\cite{5438348}, but the topology in this scenario is unpredictable due to the random links assignment scheme. Based on the NeLS constellation, the perfect match algorithm is proposed to analyze the average hop count and terminal utilization\cite{liu2016wavelength,liu2014perfect}. The wavelength convert added OSNs are proposed and analyzed in two types of single-layer LEO constellations\cite{8954693}, but the ISLs are connected randomly. Despite the LEO constellation, the ring topology of the OSN is proposed and analyzed in a GEO constellation\cite{deng2017algorithm}. The ISLs assignment scenario fitting the numerous dynamically changing ISLs and efficiently routing traffic requests in the DWROSN has not been comprehensively considered. Moreover, most LASs are concerned with establishing stable ISLs, but the possibility of using LAS to reduce on-board resource consumption has not been considered.
	
	Focusing on the above topics, the contribution of this paper is four-fold: i) We have created a dual-layer satellite constellation, and the mathematical model has been proposed. At the same time, we have proposed an additional restriction to optimize the potential ISLs set; ii) We have proposed the potential edges importance matrix (PEIM) algorithm to select the potential ISLs, and we have also proposed the links assignment scheme based on the PEIM algorithm to build the topology of the ISLs for the DWROSN; iii) We have simulated and analyzed the LAS based on the PEIM algorithm over the dual-layer constellation, and compared with the random and Greedy algorithms in terms of terminal utilization, node-to-node distance, node pair connectivity, wavelength demand, and transmission delay. iv) We have simulated and analyzed the impact of the node degree on the GEO layer satellites.
	
	The remainder of this paper is organized as follows. Section II introduces the model of the dual-layer constellation. Section III describes the PEIM algorithm and the LAS based on the PEIM algorithm. Section IV gives the simulation result and analysis. Conclusions are given in Section V.
	\section{The dual-layer satellite constellation model}
\begin{figure}[htb]
		\centering
		\subfigure[]{
			\centering
			\includegraphics[width=83mm]{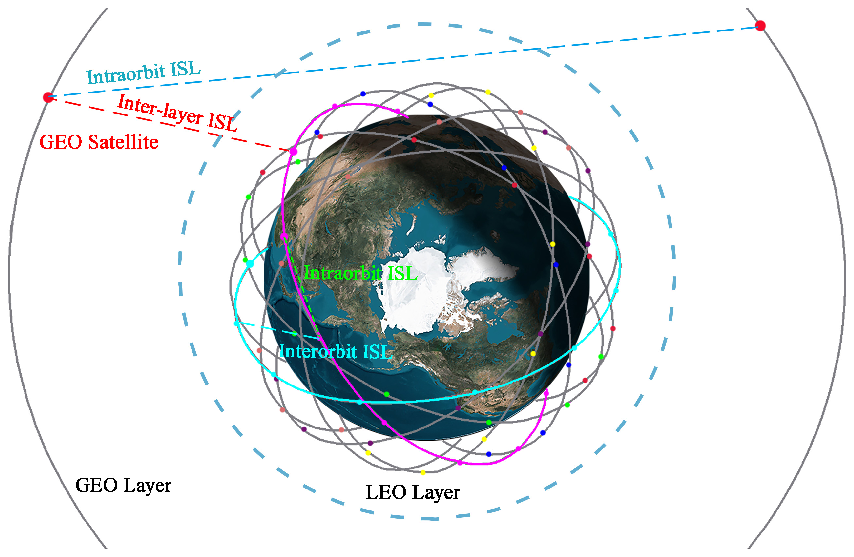}
		}
		\subfigure[]{
			\centering
			\includegraphics[width=83mm]{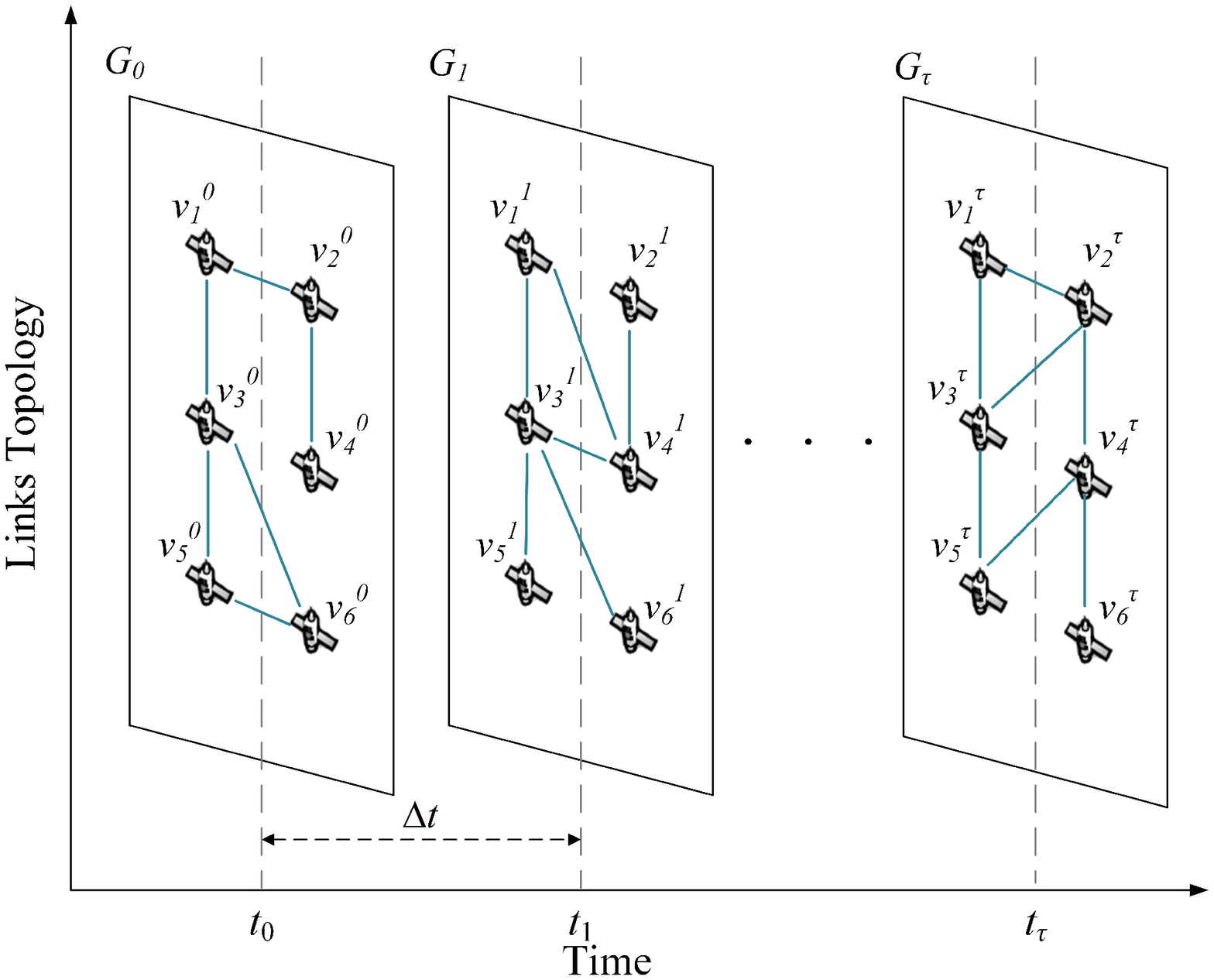}
		}
		\caption{The diagram of the double-layer OSN: (a) the constellation model and ISLs types, and (b) the snapshots for the topology of the ISLs.}
		\label{fig_1}
	\end{figure}

	The dual-layer constellation in this paper consists of two types of satellites. The diagram is shown in Fig. \ref{fig_1}(a). The constellation in this paper is based on the Walker-Delta constellation, which is described by the following five parameters: $T_l/P_l/F_l:h_l:I_l$. We use subscript $l$ to distinguish the LEO and GEO layer constellation parameters, respectively. $l=L$ represents the LEO layer constellation, and $l=G$ represents the GEO layer constellation. $T_l$ represents the total number of satellites in the constellation. $P_l$ represents the number of orbital planes. $F_l$ is the phase factor and determines the initial phase differences between adjacent satellites in adjacent orbits $\Delta_l {\omega_(l,f)} = 2\pi F_l/T_l$. $h_l$ is the orbital height (in km), and $I_l$ represents the orbital inclination (in degree). So the LEO layer constellation is ${T_L}/{P_L}/{F_L}:{h_L}:{I_L}$ is $120/10/1:1200:55$, and the GEO layer constellation is ${T_G}/{P_G}/{F_G}:{h_G}:{I_G}$ is $3/1/0:35786:0$. The primary parameters of the DWROSN are shown in Table \ref{table_1}.
	\begin{table*}[htb]
		\caption{Primary parameters of the DWROSN.\label{table_1}}
		\centering
		\begin{tabular*}{500pt}{@{\extracolsep\fill}lll@{\extracolsep\fill}}
			\toprule
			\multicolumn{1}{c}{\textbf{Parameters}} & \multicolumn{1}{c}{\textbf{LEO layer}} & \multicolumn{1}{c}{\textbf{GEO layer}} \\ 
			\midrule
			Orbit inclination ($I$)             & 55\degree      & 0\degree        \\
			Orbit altitude ($h$)                & 1200 km & 35786 km \\
			Orbit period ($U$)    &    6565 s   &  86400 s   \\
			Number of orbits ($P$)              & 10      & 1        \\
			Number of satellites per orbit ($M$) & 12      & 3        \\
			LCTs per satellite            & 5       & $5\sim8$      \\
			Inter plane spacing ($F$)           & 1       & -       \\
			\bottomrule
		\end{tabular*}
	\end{table*}
	
	As a network node, each satellite in the DWROSN has two roles. The first is an interface of traffic from terrain users through uplinks/downlinks, which is outside the scope of this paper. The second is relay services for multi-hop traffic between each source-destination satellite pair. These tasks are supported by on-board optical communication devices. To accomplish the multi-hop wavelength routing tasks, the optical elements of a satellite should compose of several LCTs, end units, and the wavelength routing unit. LCTs are equipped with the pointing acquisition and tracking (PAT) system to establish ISLs between different satellites, and each ISL supports WDM transmission tasks. End units serve the source and sink traffic tasks of the satellite. The wavelength routing unit supports routing each wavelength channel traffic from an input LCT to any output LCT. To avoid blocking, the traffic from different input LCTs can be routed into a common output LCT simultaneously only when these traffics are assigned different wavelength channels. Each ISL occupies one of the LCTs on the start and end satellite, respectively. We consider that only one ISL can be established between each satellite pair. Once an ISL has been established, both the transit traffic and the source-destination traffic can be carried through the light path unless the wavelength channel is fully loaded. Obviously, the optical ISLs as the network physical links are bidirectional. While the destination satellite is one ISL away from the source satellite, a consecutive light path with the same wavelength channel through several intermediate satellites is arranged, and we call this source-destination satellite-pair connected. 
	
	In the construction of the OSN, the usual practice is to select and establish ISLs from the potential ISLs set. The visibility, line-of-sight constraint, is the most considered condition for identifying the potential ISLs set. However, the relative position of satellites changes dynamically, so the visibility is impermanent for most of the satellite pairs. Under the Earth's Inertial Coordinate System (ECI), the position of the $m$-th satellite in the $p$-th orbit of $l$ layer constellation at time $t$ is ${S_{l,pm}}(t)$, ${S_{l,pm}}(t) = \left( {{x_{l,pm}}(t),{y_{l,pm}}(t),{z_{l,pm}}(t)} \right)$, which can be obtained as:
	\begin{align}
		\label{equ_1} {x_{l,pm}}(t) = & - R_l\cos I_l \sin \left( {\frac{{2\pi p_l}}{P_l}} \right)\sin \left[ {\omega_l t + 2\pi \left( {\frac{m_l}{M_l} + \frac{{{p_l}{F_l}}}{{{P_l}{M_l}}}} \right)} \right] + R_l\cos \left( {\frac{{2\pi p_l}}{P_l}} \right)\cos \left[ {\omega_l t + 2\pi \left( {\frac{m_l}{M_l} + \frac{{{p_l}{F_l}}}{{{P-l}{M_l}}}} \right)} \right]\\
		\label{equ_2} {y_{l,pm}}(t) = & R_l\cos I_l \cos \left( {\frac{{2\pi p_l}}{P_l}} \right)\sin \left[ {\omega_l t + 2\pi \left( {\frac{m_l}{M_l} + \frac{{{p_l}{F_l}}}{{{P_l}{M_l}}}} \right)} \right] + R_l\sin \left( {\frac{{2\pi p_l}}{P_l}} \right)\cos \left[ {\omega_l t + 2\pi \left( {\frac{m_l}{M_l} + \frac{{{p_l}{F_l}}}{{{P_l}{M_l}}}} \right)} \right]\\
		\label{equ_3} {z_{l,pm}}(t) = & R_l\sin I_l \sin \left[ {\omega_l t + 2\pi \left( {\frac{m_l}{M_l} + \frac{{{p_l}{F_l}}}{{{P_l}{M_l}}}} \right)} \right]
	\end{align}
	where $p_l,m_l \in \mathbb{Z}$, $0 \leqslant p_l \leqslant P_l-1$, $0 \leqslant m_l \leqslant M_l-1$. $R_l$ represents the orbit radius, and $\omega_l$ is the orbital angular velocity of the satellite. $R_l=R_E+h_l$, and $R_E$ is the radius of the Earth.
	
	According to the coordinates given above, the position of the satellites and the visibility of ISLs can be determined. To avoid turbulence, we stipulate that the whole light path of ISL should be 100 km above the ground. The satellite nodes and ISLs constitute the topology of the OSN. We suppose that once the satellite pair enter the visible area, the ISL can be established, and the ISL can be maintained until the satellite leaves the visible area. Obviously, the topology changes with the establishment and disconnection of ISL, and this topology is time-varying. To solve the problem of dynamic topology change in LAS design, in our model, the entire operation time is divided into a series of time slots. As shown in Fig. \ref{fig_1}(b), a sequence of undirected graph $\left\{ {{G_t}\left( {{V^t},{E^t}} \right)\left| {t \in \mathbb{N}_{+}} \right.} \right\}$ is given to capture the topology of the DWROSN. $\left\{ {{v_i}^t \in {V^t}\left| {t \in \mathbb{N}_{+},i = 1,2, \ldots ,\rm{N}} \right.} \right\}$ and $\left\{ {{e_{ij}}^t \in {E^t}\left| {t \in \mathbb{N}_{+},i = 1,2, \ldots ,\rm{N},j = 1,2, \ldots ,\rm{N}} \right.} \right\}$ represent the satellite nodes and established ISLs of the DWROSN, respectively. $\rm{N}$ is the total number of satellites. The time interval is $\Delta t$. ${G_t}$ captures the topology over $\left[ {t,t + \Delta t} \right)$. The ISLs ${{e_{ij}}^t}$ maintain connected during the time interval, so the topology is temporarily unchangeable.
	\section{Links assignment scheme based on PEIM algorithm}
	Since the visibility of the node pair can be predicted, the links topology of the DWROSN can be designed in advance by the LAS algorithm, and the establishment or disconnection of ISLs can be scheduled by the algorithm and controlled by the ground management center. The PEIM algorithm uses matrices to capture the connections of satellite nodes, the condition of LCTs, and the importance of potential ISLs to assist in the selection of the optimal ISL. By comparing the improvement in the terms of node-to-node distance and additional routing paths of each potential ISL, the optimal ISL is assigned to satellite nodes step by step. The LAS based on the PEIM algorithm diminishes the long-hops route path, reduces the average node-to-node distance, and saves the wavelength demand of the DWROSN. The construction process of the topology of the ISLs $G_t$ using LAS  based on the PEIM algorithm over $\left[ {t,t + \Delta t} \right)$ is performed as follows:
	\begin{description}
		\item[Step 1]Collect the status of OSN over $\left[ {t,t + \Delta t} \right)$. We first build satellite nodes set $\left\{ {\left. V \right|{v_i},i = 1,2, \ldots ,{\rm{N}}} \right\}$, where $\rm{N}$ is the number of satellite nodes. The node degree set is defined as $\left\{ {\left. D \right|d\left( {{v_i}} \right),{v_i} \in V} \right\}$, where $d\left( {{v_i}} \right)$ represents the number of LCTs on $v_i$ node. Record the number of generated undirected graphs with parameter $r$, and $r = 0$.
		\item[Step 2]According to the position of the satellites, calculate the visibility of each satellite node pair over $\left[ {t,t + \Delta t} \right)$. Then build potential edges matrix $\left\{ {\left. \bm{L} \right|{l_{ij}},i = 1,2, \ldots ,{\rm{N}},j = 1,2, \ldots ,{\rm{N}}} \right\}$, where $l_{ij}$ represents the relationship of visibility between satellite node pair $v_i$ and $v_j$. $\bm{L}$ is a symmetric Boole matrix and the diagonal elements are zero. If the satellite node pair could maintain visibility during $\left[ {t,t + \Delta t} \right)$, the numerical value of corresponding $l_{ij}$ is 1. Otherwise, $l_{ij} = 0$. And we build a matrix $\left\{ {\left. {{\bm{E}^*}} \right|{e_{ij}}^*,i = 1,2, \ldots ,{\rm{N}},j = 1,2, \ldots .{\rm{N}}} \right\}$ represents the ISLs decided to be established by the PEIM algorithm. ${e_{ij}}^*$ represents the ISL between $v_i$ and $v_j$. $\bm{E}^*$ is a symmetric Boole matrix, and the diagonal elements are 0. If potential ISL $l_{ij}$ is selected by the PEIM algorithm and assigned to the OSN, the numerical value of corresponding ${e_{ij}}^*$ is 1, otherwise ${e_{ij}}^* = 0$. The undirected graph of the OSN is denoted as ${G^*}\left( {V,{\bm{E}^*}} \right)$. We use $f\left( {{v_i}} \right)$ represents the number of ISLs connected with $v_i$, $f\left( {{v_i}} \right) = \sum\limits_{k = 1}^{\rm{N}} {{e_{ik}}^*}$. Obviously, $f\left( {{v_i}} \right) \leqslant d\left( {{v_i}} \right)$. In the beginning, none ISL is selected by the algorithm, so $\forall {e_{ij}}^* = 0$ and $\forall f\left( {{v_i}} \right) = 0$. 
		\item[Step 3]Calculate the importance coefficient of each potential ISL. When an ISL is established, the corresponding satellite node pair can directly transit traffic, and other multi-hops traffics can also be carried through this ISL. Therefore, establishing an ISL has an impact on the decision of routing and wavelength assignment of the traffic on OSN. The first importance coefficient matrix of the potential ISLs $\left\{ {\left. \bm{A} \right|{a_{ij}},i = 1,2, \ldots ,{\rm{N}},j = 1,2, \ldots ,{\rm{N}}} \right\}$ is defined as :
		\begin{equation}
			{a_{ij}} = \sum\limits_{{v_k},{v_n} \in V} {\delta \left( {\bm{hops}\left( {{v_k},{v_n}} \right)_{{v_k} \ne {v_n}}\left| {{l_{ij}}} \right.} \right)}
		\end{equation}
		where $\bm{hops}\left( {{v_k},{v_n}} \right)_{{v_k} \ne {v_n}}$ represents the relay hop count of the shortest route path between two different nodes $v_k$ and $v_n$ before establishing $l_{ij}$. For instance, if $v_k$ and $v_n$ can transit traffic directly, $hops\left( {{v_k},{v_n}} \right)_{{v_k} \ne {v_n}} = 1$. If $v_k$ and $v_n$ can transit traffic through one intermediate node, $hops\left( {{v_k},{v_n}} \right)_{{v_k} \ne {v_n}} = 2$. If $v_k$ and $v_n$ can not transit traffic directly or through multi-hops service, $hops\left( {{v_k},{v_n}} \right)_{{v_k} \ne {v_n}} = \inf$. ${\delta \left( {\bm{hops}\left( {{v_k},{v_n}} \right)_{{v_k} \ne {v_n}}\left| {{l_{ij}}} \right.} \right)}$ represents the decrease of relay hop count between $v_k$ and $v_n$ after $l_{ij}$ is established. Obviously, $\bm{A}$ is a symmetric matrix and the diagonal elements are 0. If potential ISL $l_{ij}$ does not exist, the value of the corresponding element $a_{ij}$ is 0.
		
		The second importance coefficient matrix of the potential ISLs $\left\{ {\left. \bm{B} \right|{b_{ij}},i = 1,2, \ldots ,{\rm{N}},j = 1,2, \ldots ,{\rm{N}}} \right\}$ is defined as :
		\begin{equation}
			{b_{ij}} = \sum\limits_{{v_k},{v_n} \in V} {\sigma \left( {\bm{paths}\left( {{v_k},{v_n}} \right)_{{v_k} \ne {v_n}}\left| {{l_{ij}}} \right.} \right)}
		\end{equation}
		where $\bm{paths}\left( {{v_k},{v_n}} \right)_{{v_k} \ne {v_n}}$ represents the optimal route paths set between two different nodes $v_k$ and $v_n$ before establishing $l_{ij}$. The optimal route paths are decided by the Dijkstra algorithm. $\sigma \left( {\bm{paths}\left( {{v_k},{v_n}} \right)_{{v_k} \ne {v_n}}\left| {{l_{ij}}} \right.} \right)$ represents the number of additional optimal route paths between $v_k$ and $v_n$ after $l_{ij}$ is established. The relay distance of these additional route paths is the same as $\bm{paths}\left( {{v_k},{v_n}} \right)_{{v_k} \ne {v_n}}$. $\bm{B}$ is a symmetric matrix and the diagonal elements are 0. If potential ISL $l_{ij}$ dose not exist, the value of corresponding element $b_{ij}$ is 0. ${\delta \left( {\bm{hops}\left( {{v_k},{v_n}} \right)_{{v_k} \ne {v_n}}\left| {{l_{ij}}} \right.} \right)}$ and $\sigma \left( {\bm{paths}\left( {{v_k},{v_n}} \right)_{{v_k} \ne {v_n}}\left| {{l_{ij}}} \right.} \right)$ will not exist non-zero values at the same time. When ${\delta \left( {\bm{hops}\left( {{v_k},{v_n}} \right)_{{v_k} \ne {v_n}}\left| {{l_{ij}}} \right.} \right)}$ has a non-zero value, which means that the optimal route paths between $v_k$ and $v_n$ are changed after $l_{ij}$ is established. We stipulate that in this case the value of $\sigma \left( {\bm{paths}\left( {{v_k},{v_n}} \right)_{{v_k} \ne {v_n}}\left| {{l_{ij}}} \right.} \right)$ is 0.
		\item[Step 4]Calculate the potential edges importance matrix $\left\{ {\left. \bm{C} \right|{c_{ij}},i = 1,2, \ldots ,{\rm{N,}}j = 1,2, \ldots ,{\rm{N,i}} \ne {\rm{j}}} \right\}$. The element of the matrix is defined as :
		\begin{equation}
			{c_{ij}} = \sum {\frac{{{a_{ij}}}}{{\max \left( \bm{A} \right)}} + \frac{{{b_{ij}}}}{{\max \left( \bm{B} \right)}}}
		\end{equation}
		where $\max \left( \bm{A} \right)$ and $\max \left( \bm{B} \right)$ are the maximum elements of $\bm{A}$ and $\bm{B}$, respectively. $\bm{C}$ is also a symmetric matrix, whose diagonal elements are 0. The value of potential edge importance consists of two parts, and these two parts show equal influence on edge importance. The first part represents the optimization of the optimal route paths between each node pair, which decrease the node-to-node distance. The second part represents the optimization of the optional optimal route paths, which brings more optional alternatives to the RWA process.
		\item[Step 5]Find the maximum value $\max \bm{C}$ of potential edges matrix. Build a temporary potential ISLs set $\left\{ {{L^*}\left| {{l_{ij}}^*} \right.} \right\}$. The elements ${{l_{ij}}^*}$ represent the potential ISLs $l_{ij}$, whose corresponding potential edges importance $c_{ij}$ is equal to $\max \bm{C}$.Then calculate the potential ISL visibility coefficient $o_{ij}$ of each ${l_{ij}}^*$. the potential ISL visibility coefficient (IVC) is defined as :
		\begin{equation}
			{o_{ij}} = \min \left( {{o_i}^*,{o_j}^*} \right)
		\end{equation}
		where $\min \left( {{o_i}^*,{o_j}^*} \right)$ is the minimal value between ${{o_i}^*}$ and ${{o_j}^*}$, which are node visibility coefficient (NVC) of $v_i$ and $v_j$, respectively. The NVC of node $v_i$ and $v_j$ is defined as:
		\begin{equation}
			{o_i}^* = \sum\limits_{k = 1}^{\rm{N}} {{l_{ik}}}
		\end{equation}
		\begin{equation}
			{o_j}^* = \sum\limits_{k = 1}^{\rm{N}} {{l_{jk}}}
		\end{equation}
		\item[Step 7]Delete ${l_{ij}}^*$ from $L^*$, whose IVC is not the minimum value. Then randomly select a potential ISL ${l_{ij}}^*$ in $L^*$, and establish this ISL into the OSN. After that, update the corresponding value of $\bm{E}^*$ and $\bm{L}$. So we set $e_{ij}^* = 1$ and $l_{ij} = 0$. 
		\item[Step 8]Check the usage of LCTs. If there is no idle LCT on $v_i$, $v_i$ cannot establish ISL with other nodes. The potential ISLs related to $v_i$ are useless in the latter process of the algorithm. So if the value of $f\left( {{v_i}} \right) $ is equal to $d\left( {{v_i}} \right)$, we set the numerical value of corresponding potential ISLs $\left\{ {{l_{ik}} = 0\left| {k = 1,2, \ldots ,{\rm{N}}} \right.} \right\}$.
		\item[Step 9]Repeat steps 3 to 8 until the numerical value of all elements of $\bm{L}$ are 0. When all potential ISLs are used, we obtain the topology of the ISLs, which is denoted as an undirected graph ${G^*}\left( {V,{\bm{E}^*}} \right)$. Then check the connectivity of this undirected graph. The sufficient condition of a connected undirected graph is that all node pairs are connectable, which denotes as $\left\{ {\forall {v_i},{v_j} \in V,paths\left( {{v_i},{v_j}} \right) \ne \emptyset } \right\}$. If ${G^*}\left( {V,{\bm{E}^*}} \right)$ is connected, $r=r+1$, and record this undirected graph as ${G_r}^*\left( {V,{\bm{E}_r}^*} \right)$. Otherwise, discard ${G^*}\left( {V,{\bm{E}^*}} \right)$, and re-generate the topology of the ISLs from step 2.
		\item[Step 10]Repeat step 2 to 9. until ${G_r}^*\left( {V,{E_r}^*} \right)$ achieve the desired amount. Then calculate the average node-to-node distance of each ${G_r}^*\left( {V,{\bm{E}_r}^*} \right)$. The average node-to-node distance is defined as :
		\begin{equation}
			{\overline H _r} = {\left. {\frac{{\sum\limits_{{v_i},{v_j} \in V} {\bm{hops}{{\left( {{v_i},{v_j}} \right)}_{{v_i} \ne {v_j}}}} }}{{{\rm{N}}\left( {{\rm{N}} - 1} \right)}}} \right|_{{G_r}^*\left( {V,{\bm{E}_r}^*} \right)}}
		\end{equation}
		where the relay hop count of the shortest paths $\bm{hops}{{\left( {{v_i},{v_j}} \right)}_{{v_i} \ne {v_j}}}$ are based on ${G_r}^*\left( {V,{E_r}^*} \right)$.
		\item[Step 11]Select ${G_r}^*\left( {V,{\bm{E}_r}^*} \right)$, whose ${\overline H _r}$ is maximum, and assign this topology as ${G^t}$. Then we have ${G^t}\left( {{V^t},{\bm{E}^t}} \right) = {G_r}^*\left( {V,{\bm{E}_r}^*} \right)$, and $\overline H  = {\overline H _r}$.
	\end{description}
	The pseudocode of the LAS based on the PEIM algorithm is shown in Algorithm \ref{algorithm_1}. The parameter $count$ denotes the amount that ${G_r}^*\left( {V,{\bm{E}_r}^*} \right)$ be generated.
	\begin{algorithm}[htb]
		\caption{Pseudocode for LAS based on PEIM algorithm.}\label{algorithm_1}
		\textbf{Input:} Constellation status, $t$, $\Delta t$, $count$. \\
		\textbf{Output:} ${G^t}\left( {{V^t},{\bm{E}^t}} \right)$, $\overline H$.\\
		\textbf{begin}
		\begin{algorithmic}[1]
			\State $r = 0$, build $\left\{ {\left. V \right|{v_i},i = 1,2, \ldots ,{\rm{N}}} \right\}$, and $\left\{ {\left. D \right|d\left( {{v_i}} \right),{v_i} \in V} \right\}$.
			\While {$r \textless count$}
				\State Generate $\left\{ {\left. \bm{L} \right|{l_{ij}},i = 1,2, \ldots ,{\rm{N}},j = 1,2, \ldots ,{\rm{N}}} \right\}$ over $\left[ {t,t + \Delta t} \right)$, and $\left\{ {\left. {{\bm{E}^*}} \right|{e_{ij}}^*,i = 1,2, \ldots ,{\rm{N}},j = 1,2, \ldots .{\rm{N}}} \right\}$.
				\While {$\exists {l_{ij}} \in \bm{L},{l_{ij}} = 1$}
					\For {$l_{ij}$ in $\bm{L}$}
						\State Calculate $a_{ij}$, and $b_{ij}$.
					\EndFor
					\State Find $\max \left( \bm{A} \right)$ and $\max \left( \bm{B} \right)$, then calculate $c_{ij}$, and build $\left\{ {{L^*}\left| {{l_{ij}}^*} \right.} \right\}$.
					\For {${l_{ij}}^*$ in $L^*$}
						\State Calculate ${o_{ij}}^*$.
					\EndFor
					\State Find $\min \left( {o_{ij}}^* \right)$, and delete the ${l_{ij}}^*$, whose ${o_{ij}}^* \textgreater \min \left( {o_{ij}}^* \right)$, from $L^*$.
					\State Randomly select ${l_{ij}}^*$ from $L^*$, and establish ISL. Then set ${e_{ij}}^* = 1$ and $l_{ij} = 0$.
					\If {$f\left( {{v_i}} \right) = d\left( {{v_i}} \right)$}
						\State set $\left\{ {{l_{ik}} = 0\left| {k = 1,2, \ldots ,{\rm{N}}} \right.} \right\}$.
					\EndIf
				\EndWhile
				\If {$\left\{ {\forall {v_i},{v_j} \in V,paths\left( {{v_i},{v_j}} \right) \ne \emptyset } \right\}$}
					\State $r++$, and denote ${G^*}\left( {V,{\bm{E}^*}} \right)$ as ${G_r}^*\left( {V,{\bm{E}_r}^*} \right)$.
				\EndIf
			\EndWhile
			\State Calculate $\overline H_r$ of each ${G_r}^*\left( {V,{\bm{E}_r}^*} \right)$. Then select ${G_r}^*\left( {V,{\bm{E}_r}^*} \right)$ with the best ${\overline H _r}$.
			\State ${G^t}\left( {{V^t},{\bm{E}^t}} \right) = {G_r}^*\left( {V,{\bm{E}_r}^*} \right)$, and $\overline H  = {\overline H _r}$.
		\end{algorithmic}
		\textbf{end}
	\end{algorithm}

	The topology of the ISLs is determined by the satellite nodes and ISLs among them, which follows the assignment scheme. There are some parameters defined to understand the features of the OSN more clearly. The parameter $\overline \alpha$ is defined to describe the average utilization of on-board LCTs of the OSN, which is defined as :
	\begin{equation}
		\overline \alpha  = \frac{{\sum\limits_{{v_i} \in V} {f\left( {{v_i}} \right)} }}{{\sum\limits_{{v_i} \in V} {d\left( {{v_i}} \right)} }}
	\end{equation}

	The parameter $\beta$ is defined to describe the satellite node pair connectivity, which is defined as :
	\begin{equation}
		\beta  = \frac{{\sum\limits_{{v_i},{v_j} \in V} {{{\left. {\bm{con}{{\left( {{v_i},{v_j}} \right)}_{{v_i} \ne {v_j}}}} \right|}_{{{\left\{ {{N_{hops}}} \right\}}_{\max }}}}} }}{{{\rm{N}}\left( {{\rm{N}} - 1} \right)}}
	\end{equation}
	where ${{{\left\{ {{N_{hops}}} \right\}}_{\max }}} \in \mathbb{N}_{+}$ denotes the maximum allowable relay hop count of the route paths. ${\sum\limits_{{v_i},{v_j} \in V} {{{\left. {\bm{con}{{\left( {{v_i},{v_j}} \right)}_{{v_i} \ne {v_j}}}} \right|}_{{{\left\{ {{N_{hops}}} \right\}}_{\max }}}}} }$ counts the connectable node pairs. ${{{\left. {\bm{con}{{\left( {{v_i},{v_j}} \right)}_{{v_i} \ne {v_j}}}} \right|}_{{{\left\{ {{N_{hops}}} \right\}}_{\max }}}}}$ represents the connectivity between two different satellite nodes $v_i$ and $v_j$. If there exist at least one route path between $v_i$ and $v_j$, whose relay hop count is equal or less than ${{{\left\{ {{N_{hops}}} \right\}}_{\max }}}$, this node pair is connectable.
	
	In wavelength routing OSN, the route paths and wavelength channels of the traffic requests are decided by the routing and wavelength assignment (RWA) algorithm. To evaluate the performance of the wavelength routing OSN under the network topology ${G^t}\left( {{V^t},{\bm{E}^t}} \right)$ generated by LAS algorithm, we assume that there is a set of traffic requests $\left\{ {\bm{Q}\left| {\bm{req}{{\left( {{v_i},{v_j}} \right)}_{{v_i} \ne {v_j}}}} \right.} \right\}$ to be accommodate. There are ${\rm{N}}\left( {{\rm{N}} - 1} \right)/2$ requests in $\bm{Q}$, and all requests ask for the same bandwidth which is the capacity of a wavelength channel. The routing and wavelength assignment (RWA) algorithm is performed as Algorithm \ref{algorithm_2}. ${N_\lambda }$ represents the wavelength demand of the OSN, ${N_\lambda } \in \mathbb{N}_{+}$. $\left\{ {{\bm{W}_\lambda }\left| {\bm{channel}\left( {{e_{ij}},\lambda } \right)} \right.} \right\}$ represents the set of $\lambda$-th wavelength channel. ${\bm{channel}\left( {{e_{ij}},\lambda } \right)}$ represents the $\lambda$-th wavelength channel of ISL $e_{ij}$. The wavelength channel continuity condition requires that the route path of traffic request ${\bm{req}{{\left( {{v_i},{v_j}} \right)}_{{v_i} \ne {v_j}}}}$ uses the same wavelength channel to transit. For example, $v_i$ and $v_j$ transit traffic request ${\bm{req}{{\left( {{v_i},{v_j}} \right)}_{{v_i} \ne {v_j}}}}$ through a intermediate node $v_k$ using $\lambda$-th wavelength channel. The wavelength channel continuity condition requires that ${\bm{channel}\left( {{e_{ik}},\lambda } \right)}$, and ${\bm{channel}\left( {{e_{kj}},\lambda } \right)}$ have enough bandwidth to transit traffic request ${\bm{req}{{\left( {{v_i},{v_j}} \right)}_{{v_i} \ne {v_j}}}}$. The transmitting delay of a route path consists of two parts. The first is the propagation delay determined by the distance of the light path for the node pair. The second is processing delay, caused by the LCTs and the wavelength routing unit.
	\begin{algorithm}[htb]
		\caption{Pseudocode for RWA algorithm.}\label{algorithm_2}
		\textbf{Input:} ${G^t}\left( {{V^t},{\bm{E}^t}} \right)$, and ${{{\left\{ {{N_{hops}}} \right\}}_{\max }}}$.\\
		\textbf{Output:} Route paths, ${N_\lambda }$, and $\beta$.\\
		\textbf{begin}
		\begin{algorithmic}[1]
			\State Generate traffic request set $\left\{ {\bm{Q}\left| {\bm{req}{{\left( {{v_i},{v_j}} \right)}_{{v_i} \ne {v_j}}}} \right.} \right\}$. Reset $N_\lambda = 1$, and generate wavelength channels set $\left\{ {{\bm{W}_\lambda }\left| {\bm{channel}\left( {{e_{ij}},\lambda } \right)} \right.} \right\}$ based on ${G^t}\left( {{V^t},{\bm{E}^t}} \right)$. $\lambda  = 1,2, \ldots ,{N_\lambda }$.
			\While {$\bm{Q} \ne \emptyset$}
				\State Randomly select ${\bm{req}{{\left( {{v_i},{v_j}} \right)}_{{v_i} \ne {v_j}}}} \in \bm{Q}$, and find optimal route paths set $\left\{ {{\bm{R}_{ij}}^*\left| {\bm{route}^*{{\left( {{v_i},{v_j}} \right)}_{{v_i} \ne {v_j}}}} \right.} \right\}$. The relay hop count of these route paths should be equal to or less than ${{{\left\{ {{N_{hops}}} \right\}}_{\max }}}$. Sort $\bm{route}^*{{\left( {{v_i},{v_j}} \right)}_{{v_i} \ne {v_j}}} \in {{R_{ij}}^*}$ in ascending order of relay hop count. Then sort the same relay hop count $\bm{route}^*{{\left( {{v_i},{v_j}} \right)}_{{v_i} \ne {v_j}}}$ in ascending order of transmitting delay.
				\If {${{\bm{R}_{ij}}^*} \ne \emptyset$}
					\For {$\bm{route}^*{{\left( {{v_i},{v_j}} \right)}_{{v_i} \ne {v_j}}} \in {{\bm{R}_{ij}}^*}$}
						\State Check the wavelength channel continuity condition from the 1-st to $\lambda$-th wavelength channel.
						\If {$\lambda$-th wavelength channel satisfy the wavelength continuity condition}
							\State Assign traffic request $\bm{req}{{\left( {{v_i},{v_j}} \right)}_{{v_i} \ne {v_j}}}$ into the OSN through route path $\bm{route}^*{{\left( {{v_i},{v_j}} \right)}_{{v_i} \ne {v_j}}}$ using $\lambda$-th wavelength channel.
							\State \textbf{break}
						\EndIf
					\EndFor
					\If {Fail to assign traffic request $\bm{req}{{\left( {{v_i},{v_j}} \right)}_{{v_i} \ne {v_j}}}$ with current wavelength channels}
						\State ${N_\lambda }++$, and assign traffic request $\bm{req}{{\left( {{v_i},{v_j}} \right)}_{{v_i} \ne {v_j}}}$ into the OSN through the shortest route path using the new wavelength channel. 
					\EndIf
				\EndIf
				\State Remove $\bm{req}{{\left( {{v_i},{v_j}} \right)}_{{v_i} \ne {v_j}}}$ from $\bm{Q}$.
			\EndWhile
		\end{algorithmic}
		\textbf{end}
	\end{algorithm}
	\section{Simulation Results and Analysis}
	To evaluate the performance of the PEIM algorithm, this paper takes a dual-layer constellation as an example to establish ISLs and build wavelength routing OSN. The parameters of the dual-layer constellation and on-board LCTs are shown in Table \ref{table_1}. To analyze the advantages brought by LAS based on the PEIM algorithm, this paper compares with OSN generated by LAS based on the arbitrary connection algorithm (ACT)\cite{5438348}, and LAS based on the Greedy algorithm\cite{344508}. Since the topology of the ISLs generated by the ACT and Greedy algorithms are random, this paper generates 100 topologies based on the ACT and Greedy algorithms in every time slot, respectively. The topologies generated by ACT and Greedy algorithms should be connectable, and we select the topology with the best average node-to-node distance to compare with the topology generated by the PEIM algorithm. The desired amount $count$ of the PEIM algorithm in Algorithm \ref{algorithm_1} Line 2 is also set to 100. The time interval affects the frequency of establishing and disconnecting the ISLs. Short time interval requires the high performance of the PAT unit on LCTs, as a result, the time interval is as long as possible. The time interval $\Delta t$ is set to 2000 s, and the entire operation time of the simulation is 0$\sim$20000 s.
	\subsection{The potential ISLs selection}
	The types of ISLs are shown in Fig.\ref{fig_1}(b). Intraorbit ISLs interconnect the satellites belonging to the same orbit, and interorbit ISLs interconnect the satellites belonging to different orbits. Inter-layer ISLs interconnect the satellites belonging to different constellation layers, and the inter-layer ISLs belong to the type of interorbit ISLs. Same-layer ISLs interconnect the satellites belong to the same constellation layer, including all intraorbit ISLs and some interorbit ISLs. We first calculate the position of all satellite nodes and the visibility of all node pairs for the DWROSNs based on Eq.\ref{equ_1} to \ref{equ_3}, and the calculation time step is 1 s.
	
\begin{figure}[htb]
		\centering
		\subfigure[]{
			\centering
			\includegraphics[width=55mm]{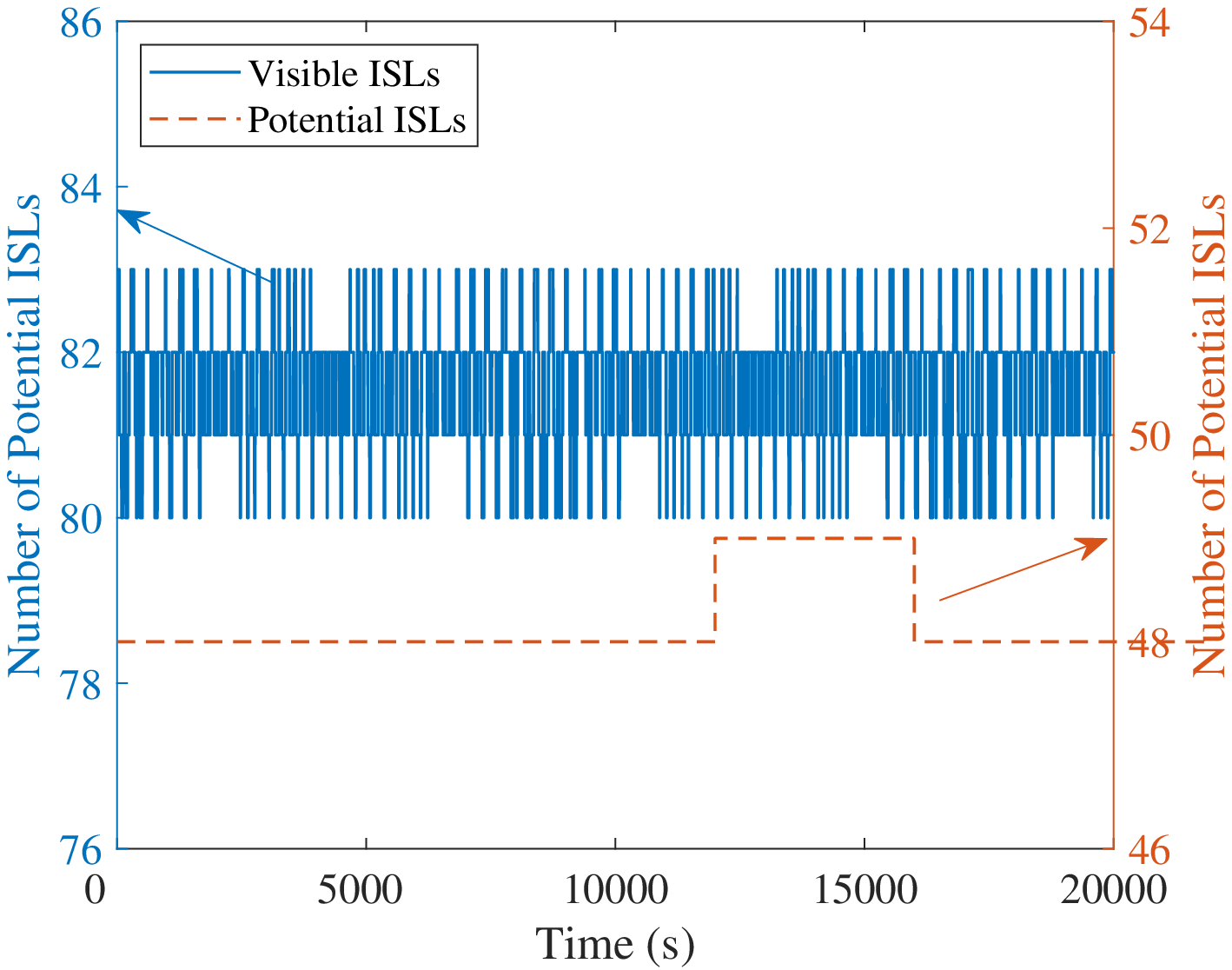}
		}
		\subfigure[]{
			\centering
			\includegraphics[width=55mm]{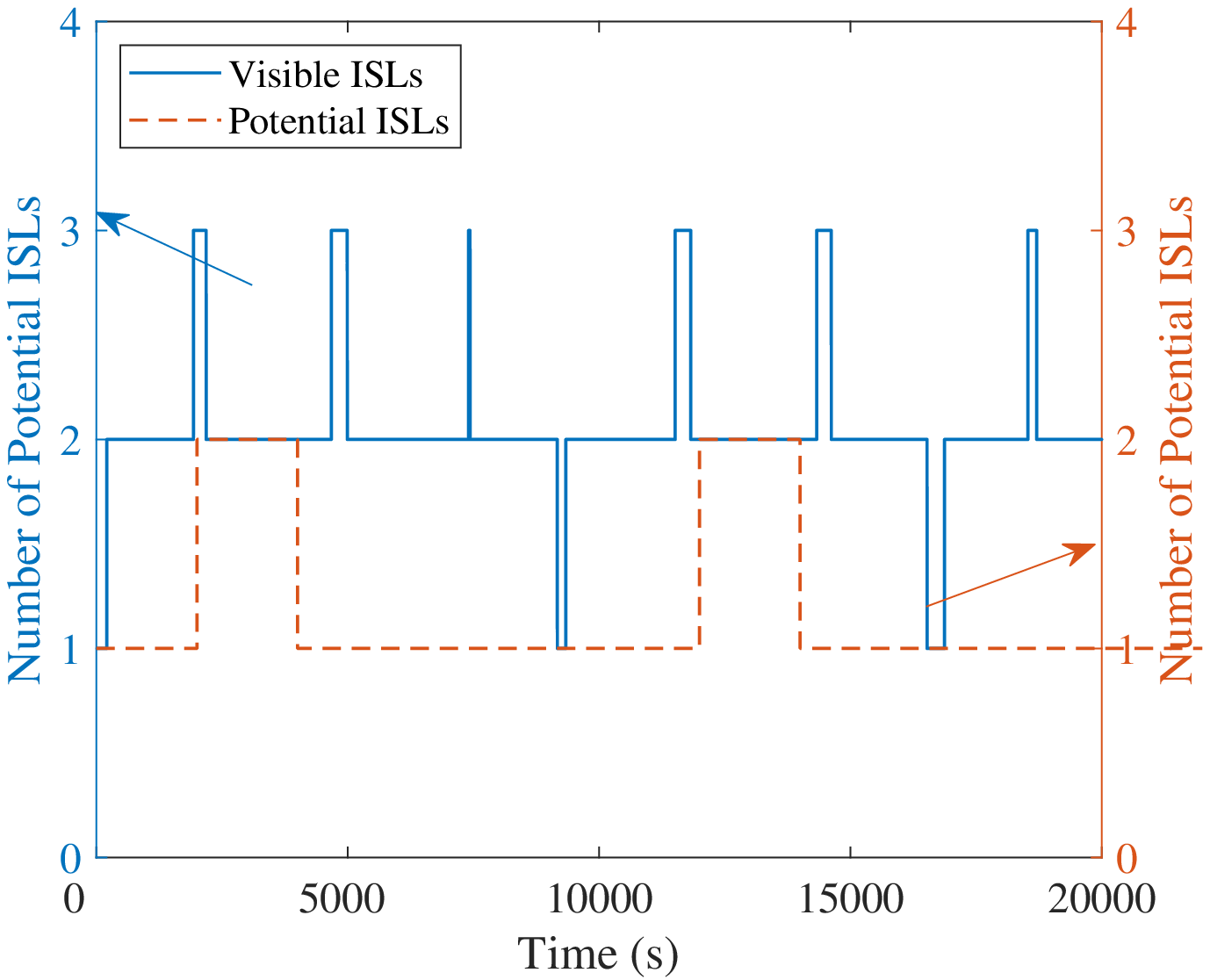}
		}
		\subfigure[]{
			\centering
			\includegraphics[width=55mm]{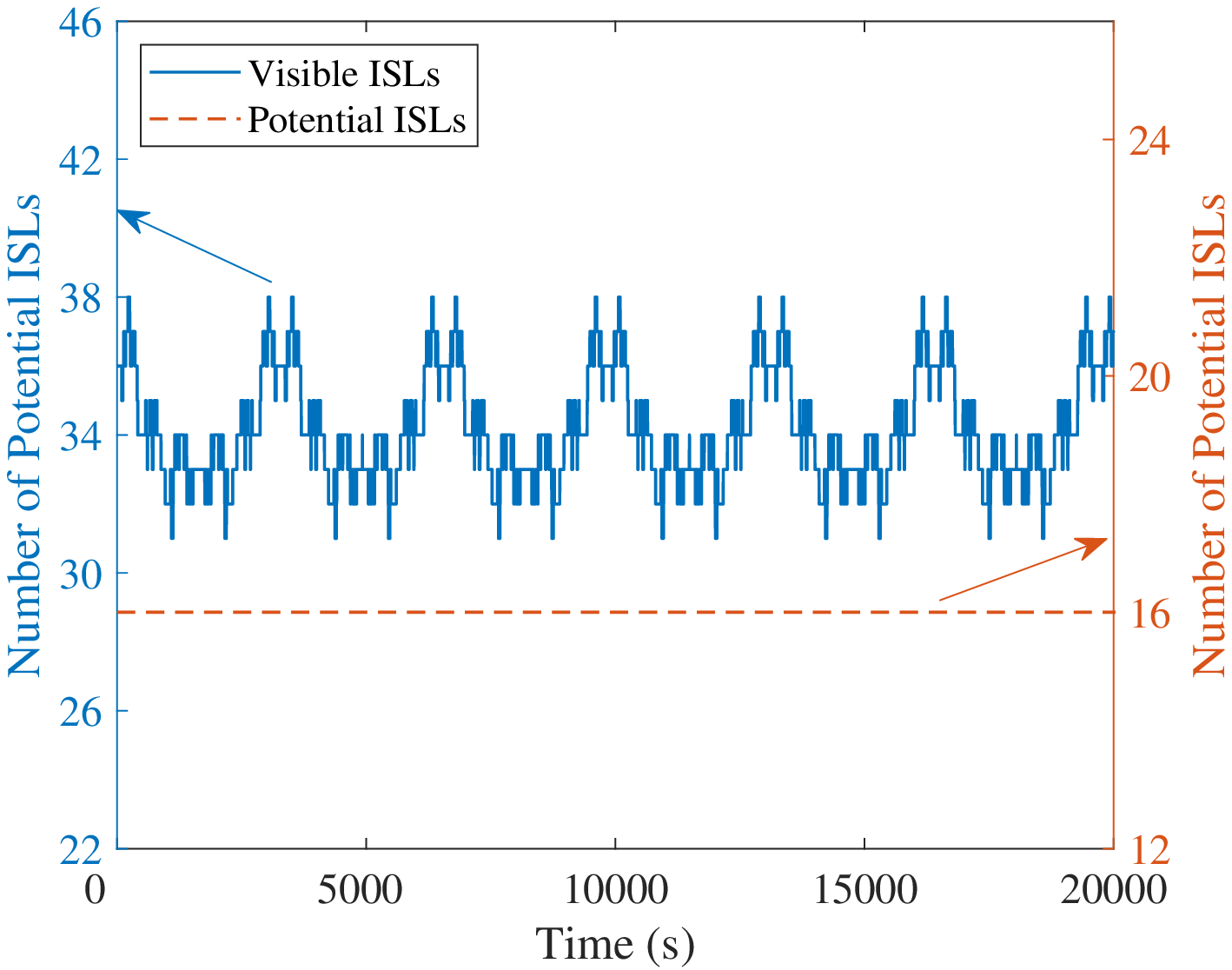}
		}
		\caption{The number of visible and potential ISLs: (a) The inter-layer ISLs of $S_{G,00}$, (b) The inter-layer ISLs of $S_{L,00}$, and (c) The same-layer ISLs of $S_{L,00}$.}
		\label{fig_2}
	\end{figure}

	Then we build potential ISLs set for each time slot based on the visibility of ISLs. All permanent ISLs and the temporary permanent same-layer ISLs are selected to build the potential ISLs set for each time slot. Fig. \ref{fig_2} shows the number of ISLs during the entire operation time. The solid lines represent the number of visible ISLs, and the dashed lines represent the number of potential ISLs. Fig. \ref{fig_2}(a) and (b) show the number of visible and potential inter-layer ISLs of a GEO satellite $S_{G,00}$ and an LEO $S_{L,00}$, respectively. The orbit altitude of GEO layer satellites is higher than LEO layer satellites, so $S_{G,00}$ has more visible and potential inter-layer ISLs than $S_{L,00}$. The number of visible inter-layer ISLs of $S_{G,00}$ and $S_{L,00}$ are 80$\sim$83 and 1$\sim$3, respectively. These visible ISLs are all impermanent. The number of potential inter-layer ISLs of $S_{G,00}$ and $S_{L,00}$ are 48$\sim$49 and 1$\sim$2, respectively. Fig. \ref{fig_2}(c) shows the number of visible and potential same-layer ISLs of $S_{L,00}$. The number of visible same-layer ISLs of $S_{L,00}$ is 31$\sim$38, and the potential same-layer ISLs for $S_{L,00}$ is 16. The potential ISLs selection eliminates the impermanent ISLs in each time slot and reduces the potential ISLs amount. For instance, there are 3900 visible ISLs in time slot $\left[ {0,2000} \right)$. Among them, 345 ISLs are inter-layer ISLs, and 3550 ISLs are same-layer ISLs. At the same time, the total number of potential ISLs is 1105. Among them, 142 ISLs are inter-layer ISLs, and 963 ISLs are same-layer ISLs. Through the process of selecting the potential ISLs, the topology of the DWROSN constructed with these potential ISLs is stable during each time slot.
	\subsection{The node-to-node distance and node pair connectivity}
	Based on the potential ISLs set in each time slot, topologies of the DWROSNs are established by the LASs based on ACT, Greedy, and PEIM algorithms. The number of on-board LCTs of LEO layer satellites is 5, denotes as $d_L=5$, and the number of on-board LCTs of GEO layer satellites is 6, denotes as $d_G=6$.
	
\begin{figure}[htb]
		\centering
		\subfigure[]{
			\centering
			\includegraphics[width=55mm]{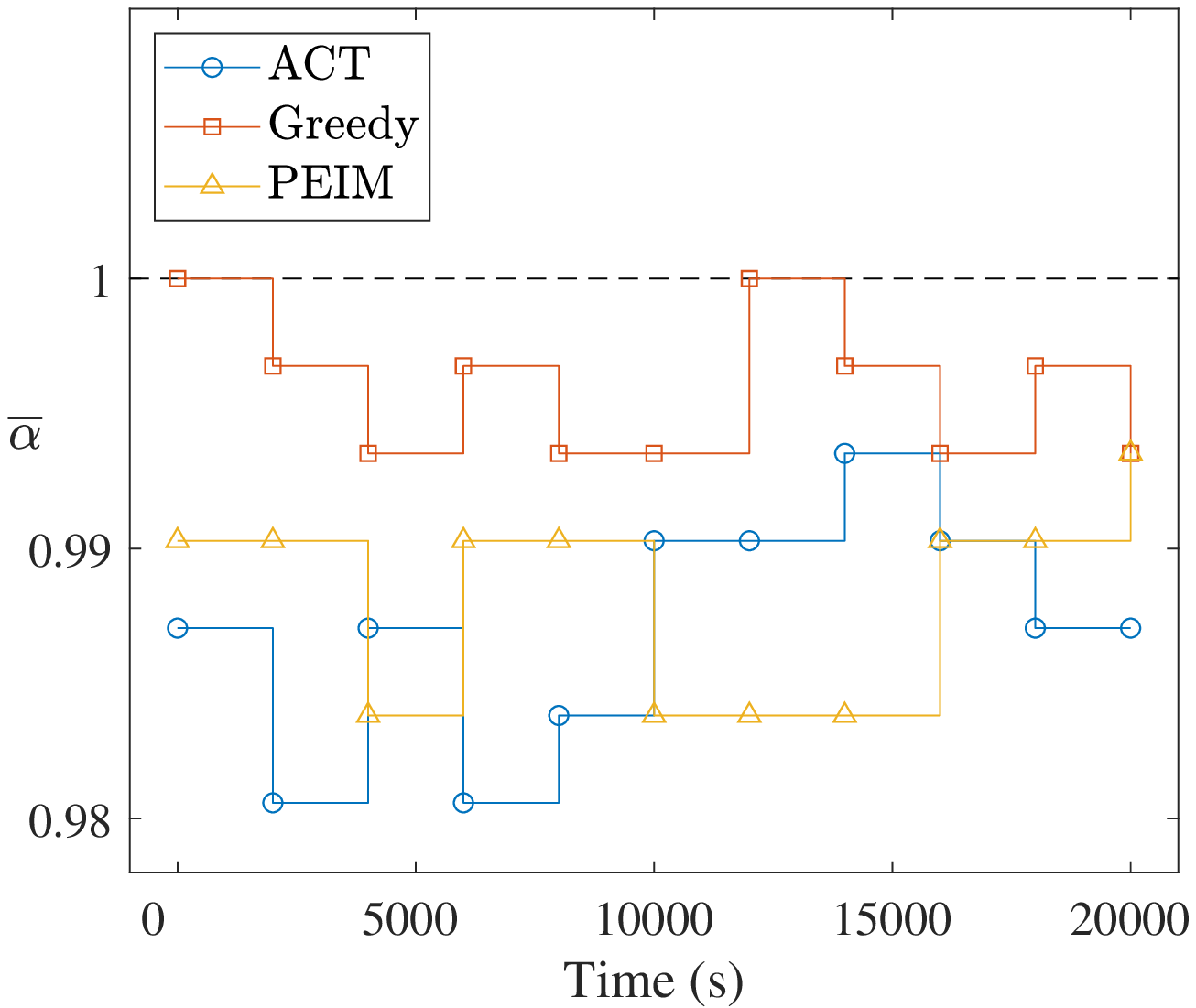}
		}
		\subfigure[]{
			\centering
			\includegraphics[width=55mm]{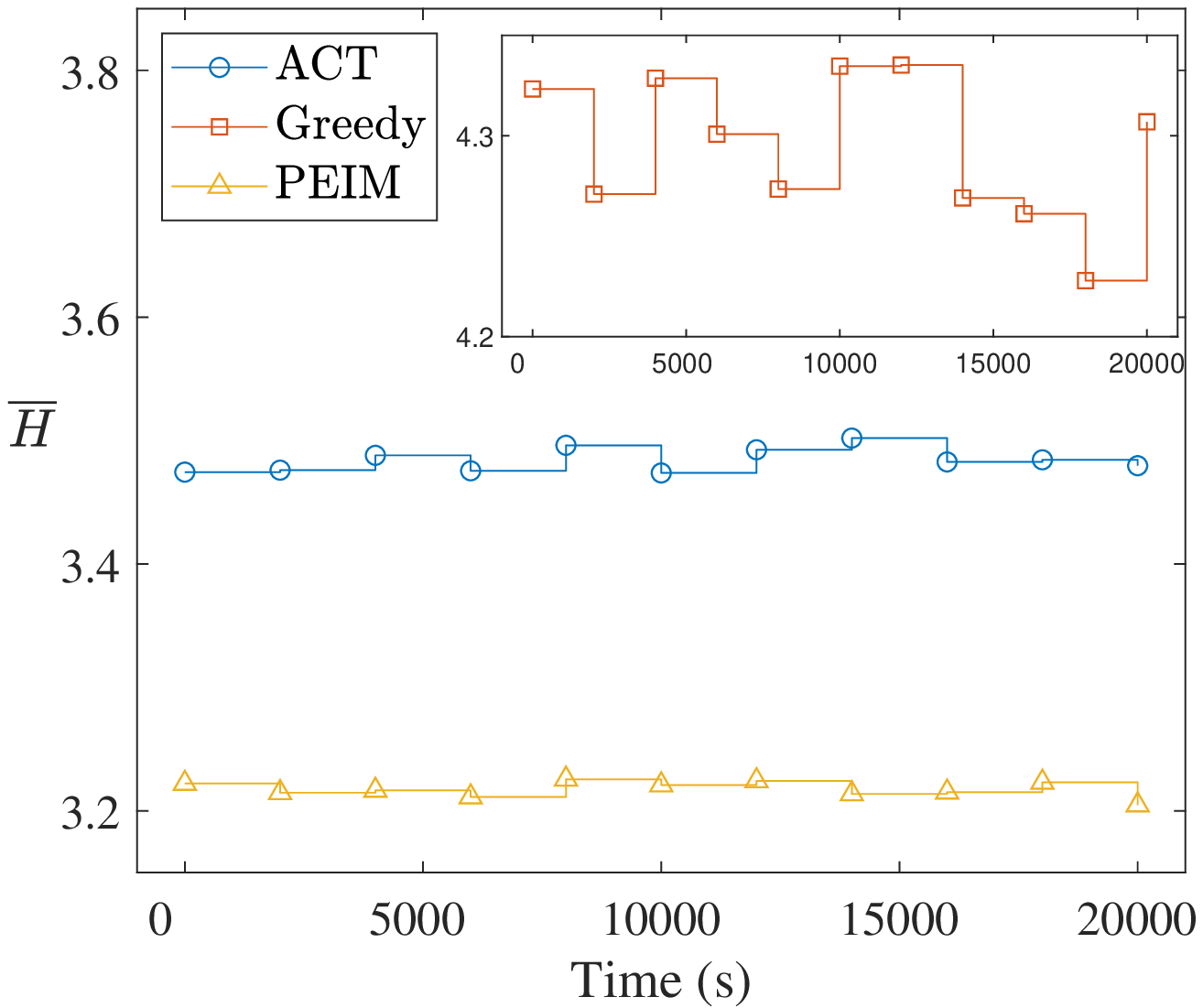}
		}
		\subfigure[]{
			\centering
			\includegraphics[width=55mm]{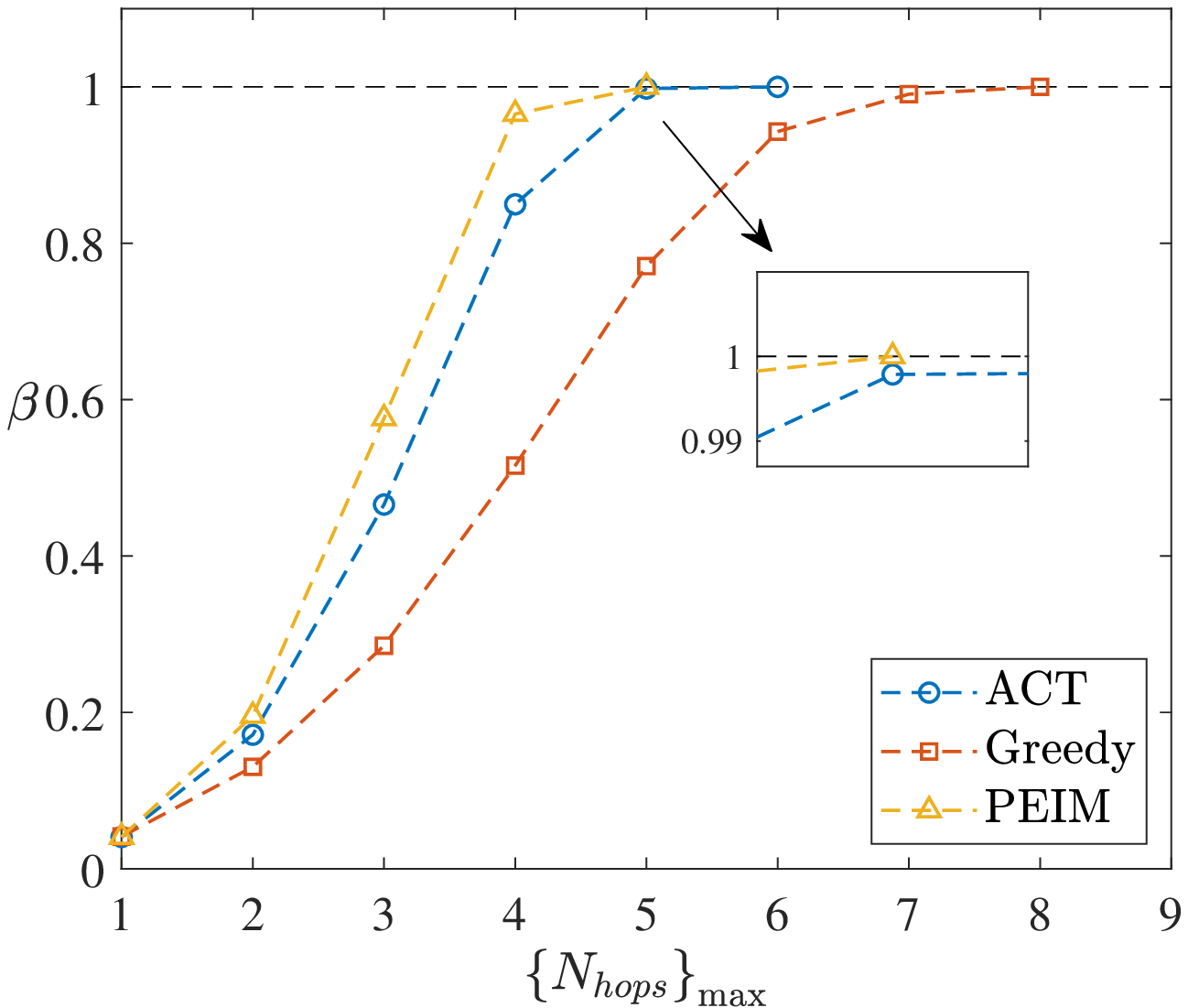}
		}
		\caption{The result of the DWROSNs generated by the algorithms: (a) the terminal utilization in each time slot, (b) the average node-to-node distance in each time slot, and (c) the relationship between the maximum allowable relay hop count and the node pair connectivity during time slot $\left[ {0,2000} \right)$.}
		\label{fig_3}
	\end{figure}

	The average terminal utilization of DWROSNs is shown in Fig. \ref{fig_3} (a). The topologies generated by the PEIM algorithm show great LCTs utilization in assigning ISLs for DWROSNs, where the average terminal utilization of the topology is higher than 98\% during the entire operation time. The mean values of $\overline \alpha$ for the topologies generated by the ACT, Greedy, and PEIM algorithms are 98.82\%, 99.59\%, and 98.71\%, respectively. The PEIM algorithm shows nearly the same terminal utilization compared with the ACT algorithm and is slightly lower than the Greedy algorithm. Fig. \ref{fig_3} (b) shows the average node-to-node distance of DWROSNs generated by the three algorithms. The topologies generated by the PEIM algorithm show the best node-to-node distance, where the mean value of $\overline H$ is 3.218 hops. The mean values of $\overline H$ for the ACT and Greedy algorithms are 3.484 hops and 4.294 hops, respectively. Compared with ACT and Greedy algorithms, the PEIM algorithm reduces the average node-to-node distance by 7.6\% and 25.1\%. 
	
	Fig \ref{fig_3} (c) shows the relationship between the maximum allowable relay hop count and the node pair connectivity during $\left[ {0,2000} \right)$. The PEIM algorithm shows a better ability in reducing the node-to-node distance. Except ${\left\{ {{N_{hops}}} \right\}_{\max }} = 1$, the DWROSN generated by the PEIM algorithm has better node pair connectivity than ACT and Greedy algorithms as the ${\left\{ {{N_{hops}}} \right\}_{\max }}$ grows. The DWROSN generated by the PEIM algorithm reaches full connectivity within 5 hops, while the ACT and Greedy algorithms require 6 hops and 8 hops, respectively. Table \ref{table_2} details the distribution probability of the relay hop count for the node pairs in the DWROSNs generated by the PEIM algorithm compared with ACT and Greedy algorithms during time slot $\left[ {0,2000} \right)$. The distribution probability of $N_{hops}$ relay hop count is defined as the ratio that the number of node pairs, whose relay hop count of the shortest paths are equal to $N_{hops}$, to all node pairs in the DWROSN. So the average node-to-node distance is also the weighted average of $N_{hops}$ with distribution probability as the weight. Higher terminal utilization represents more directly connected node pairs, so the DWROSN generated by the Greedy algorithm has a higher distribution probability for $N_{hops} = 1$ than ACT and PEIM algorithms. The PEIM algorithm reduces the long-distance relay, and about 96\% of the node pairs in the DWROSN generated by the PEIM algorithm have no more than 4 hops of the relay paths. For ACT and Greedy algorithms, there are about 15\% and 48\% of the node pairs, whose relay hop count is more than 4 hops, respectively.
	\begin{table*}[htb]
		\caption{Distribution probability of relay hop count for the node pairs in DWROSNs generated by the algorithms during time slot $\left[ {0,2000} \right)$.\label{table_2}}
		\centering
		\renewcommand{\arraystretch}{1.2}
		\begin{tabular*}{500pt}{@{\extracolsep\fill}cllllllllll@{\extracolsep\fill}}
			\toprule
			\multirow{2}{*}{\textbf{LAS Type}} &
			\multicolumn{8}{c}{\bm{${N_{hops}}$}} &
			\multicolumn{1}{c}{\multirow{2}{*}{\bm{$\overline \alpha$}}} &
			\multicolumn{1}{c}{\multirow{2}{*}{\bm{$ \overline H $}}} \\ \cline{2-9}
			&
			\multicolumn{1}{c}{\textbf{1}} &
			\multicolumn{1}{c}{\textbf{2}} &
			\multicolumn{1}{c}{\textbf{3}} &
			\multicolumn{1}{c}{\textbf{4}} &
			\multicolumn{1}{c}{\textbf{5}} &
			\multicolumn{1}{c}{\textbf{6}} &
			\multicolumn{1}{c}{\textbf{7}} &
			\multicolumn{1}{c}{\textbf{8}} &
			\multicolumn{1}{c}{} &
			\multicolumn{1}{c}{} \\ \midrule
			\textbf{ACT} & 4.07\% & 13.07\% & 29.44\% & 38.40\% & 14.81\% & 0.21\% & - & - & 98.71\% & 3.475 \\
			\textbf{Greedy} & 4.12\% & 8.92\% & 15.49\% & 23.04\% & 25.51\% & 17.21\% & 4.78\% & 0.93\% & 100\% & 4.323 \\
			\textbf{PEIM} & 4.08\% & 15.49\% & 38.04\% & 38.93\% & 3.47\% & - & - & - & 99.03\% & 3.222\\ \bottomrule
		\end{tabular*}
	\end{table*}
	\subsection{The wavelength demand and average delay}
	The traffic transit performance of the DWROSNs in each time slot is simulated. The node degree is set to $d_L = 5$ and $d_G = 6$. The traffic requests are generated and assigned to the DWROSNs following the RWA algorithm in Algorithm \ref{algorithm_2}. For the presented simulation, the processing delay is 10 ms for each hop of the route path between the node pair. The simulations are performed 10 times for the DWROSNs generated by the three algorithms in each time slot, respectively.
	
\begin{figure}[htb]
		\centering
		\subfigure[]{
			\centering
			\includegraphics[width=83mm]{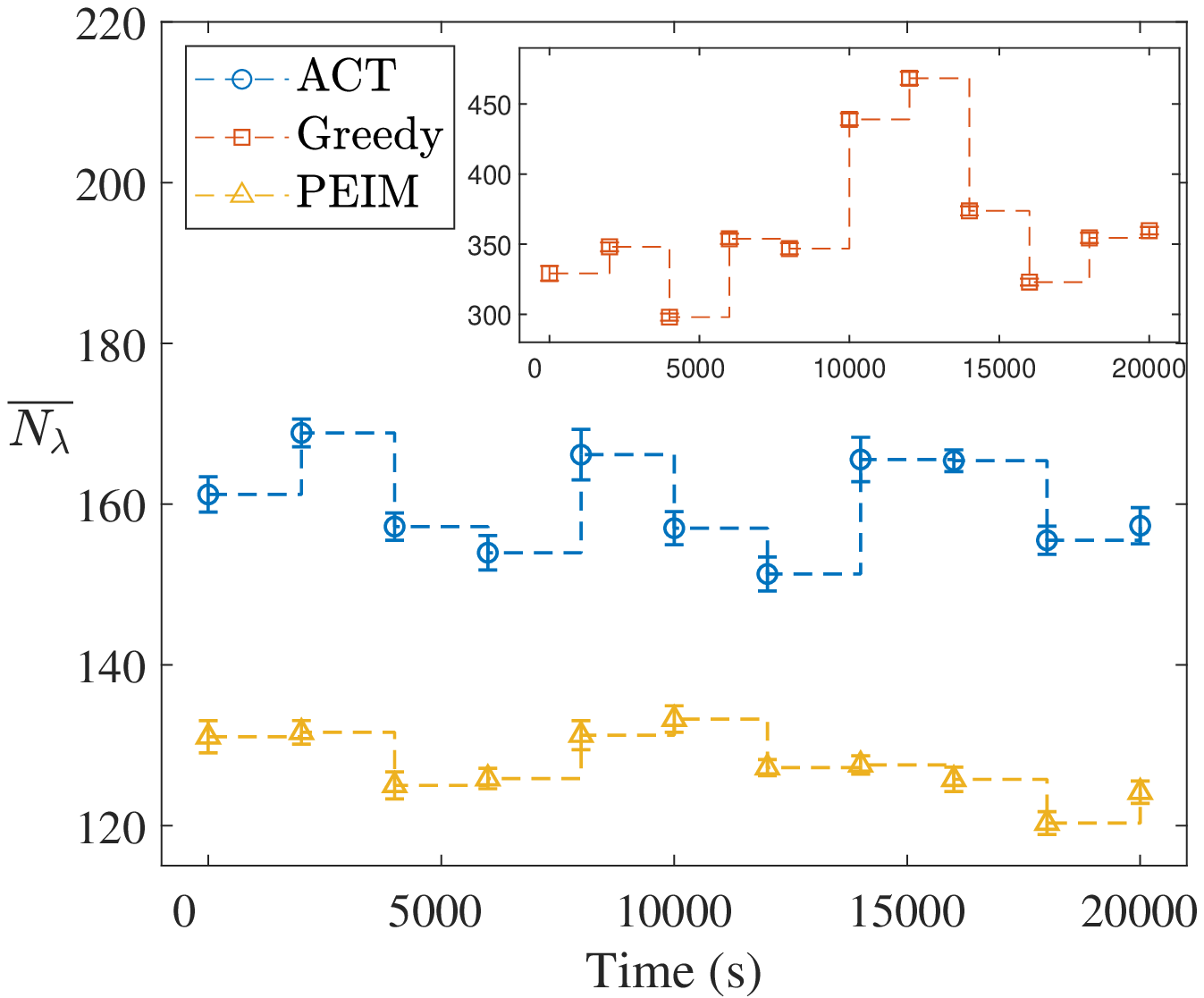}
		}
		\subfigure[]{
			\centering
			\includegraphics[width=83mm]{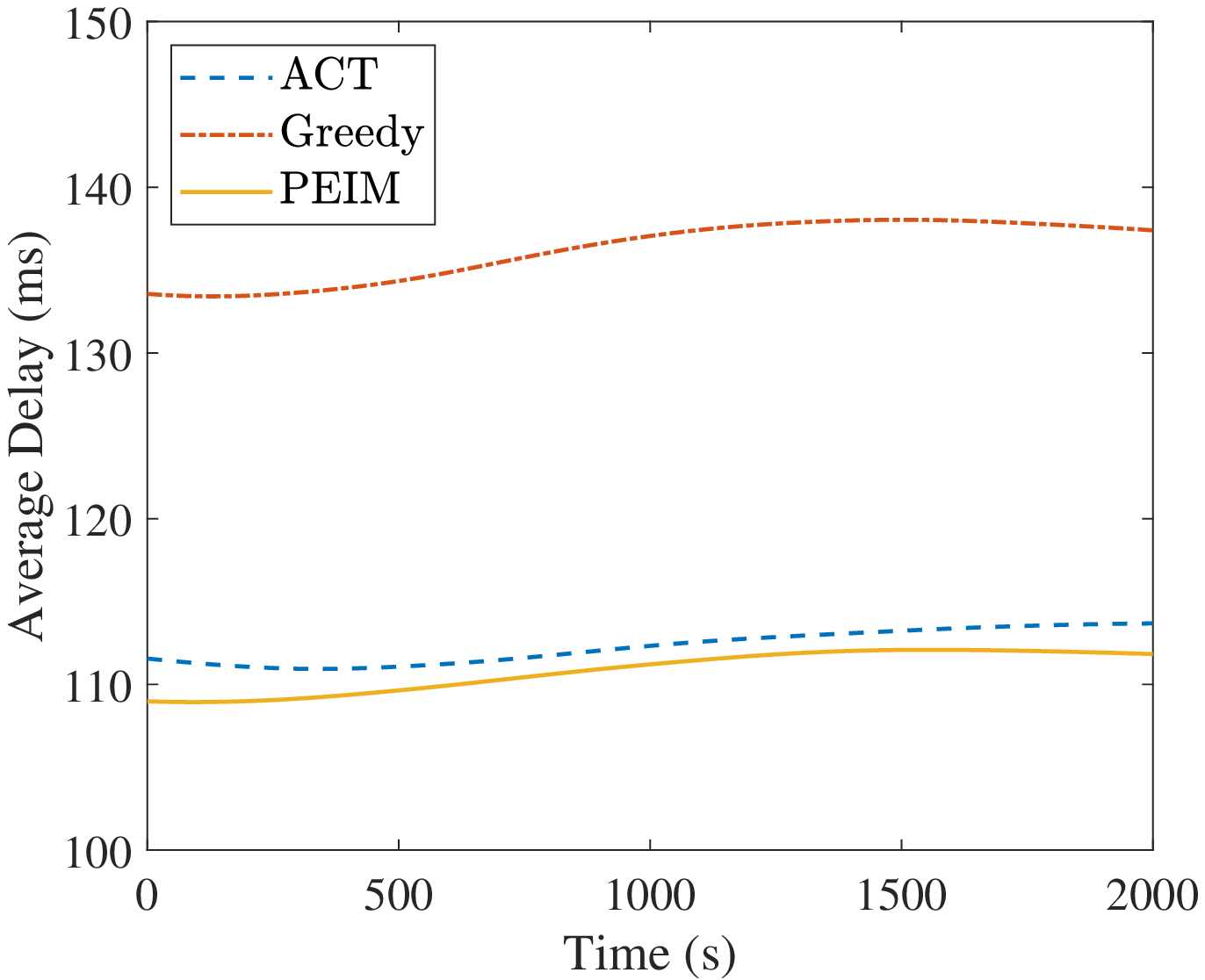}
		}
		\caption{Simulation results of traffic requests in the DWROSNs: (a) Wavelength demand in each time slot, and (b) average node pair delay during time slot $\left[ {0,2000} \right)$.}
		\label{fig_4}
	\end{figure}

	Fig. \ref{fig_4} (a) shows the wavelength demand of the DWROSNs with full node pair connectivity in each time slot. The maximum allowable relay hop count is set to infinite so that all node pairs have at least one route path to carry the traffic request following the RWA algorithm in \ref{algorithm_2}. The points in the figure represent the average value of wavelength demand that the DWROSNs reach full connectivity, and the error bars represent the maximum and minimum values of wavelength demand in the corresponding time slot, respectively. The DWROSNs generated by the PEIM algorithm require less wavelength demand to reach the full node pair connectivity. The average value of $\overline {{N_\lambda }}$ during the entire operation time for DWROSNs generated by the PEIM algorithm is 127.54, while the ACT and Greedy algorithms require 159.95 and 363.10, respectively. The PEIM algorithm saves the wavelength resources by 20.3\% and 64.9\%, compared with ACT and Greedy algorithms, respectively.

	Fig. \ref{fig_4} (b) shows the average traffic requests transmitting delay of the DWROSNs during time slot $\left[ {0,2000} \right)$. Despite the topology of the ISLs being stable during each time slot, the distances of light paths among the node pairs are dynamically changing due to the high-speed moving satellites. During time slot $\left[ {0,2000} \right)$, the traffic requests in DWROSN generated by the PEIM algorithm have shorter average transmitting delay than those in DWROSNs generated by the ACT and Greedy algorithms. The average transmitting delay of the traffic requests are 112.3 ms, 136.2 ms and 110.8 ms for the DWROSNs generated by the ACT, Greedy, and PEIM algorithms, respectively.
	\subsection{The impact of node degree in GEO layer}
\begin{figure}[htb]
		\centering
		\subfigure[]{
			\centering
			\includegraphics[width=83mm]{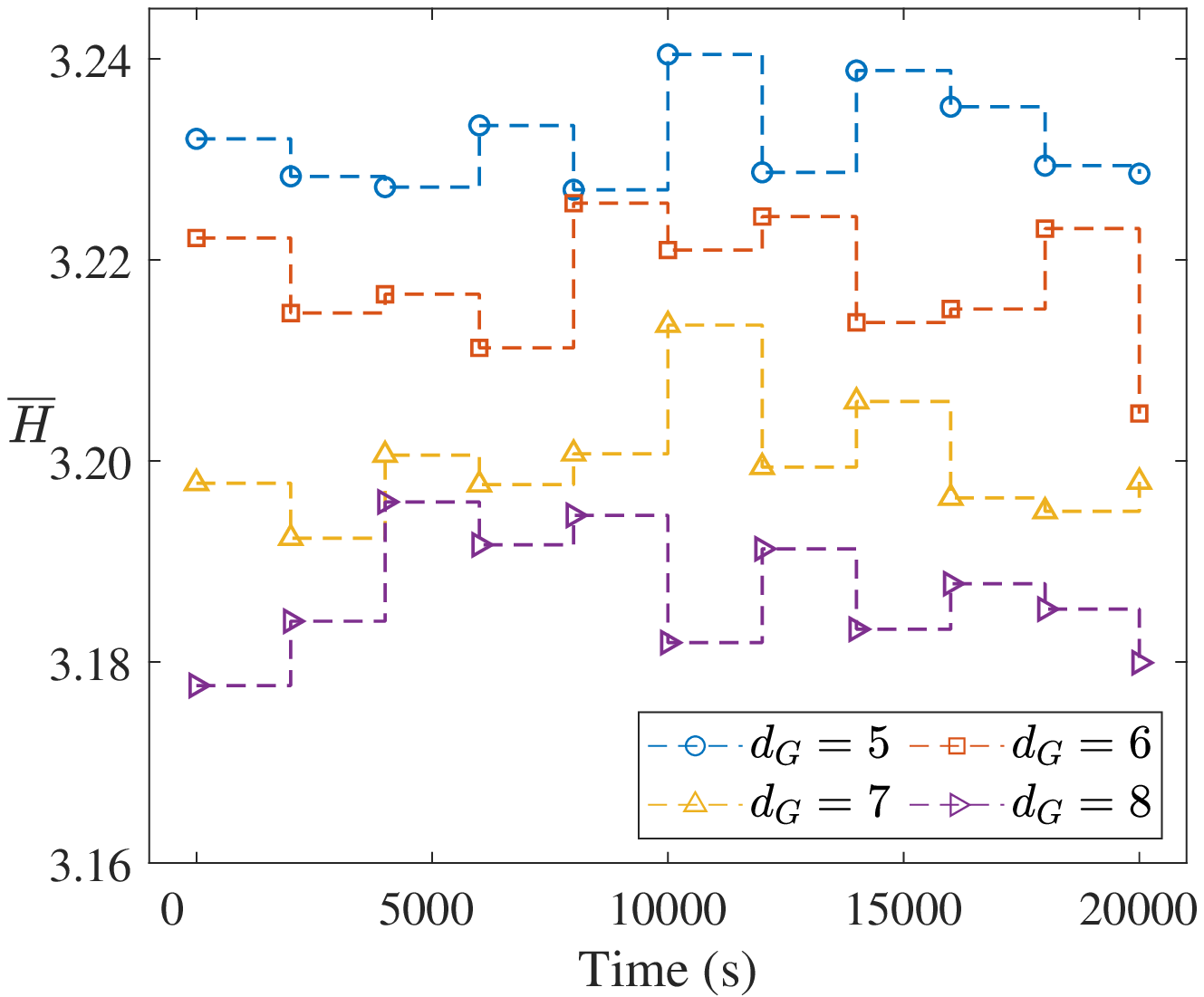}
		}
		\subfigure[]{
			\centering
			\includegraphics[width=83mm]{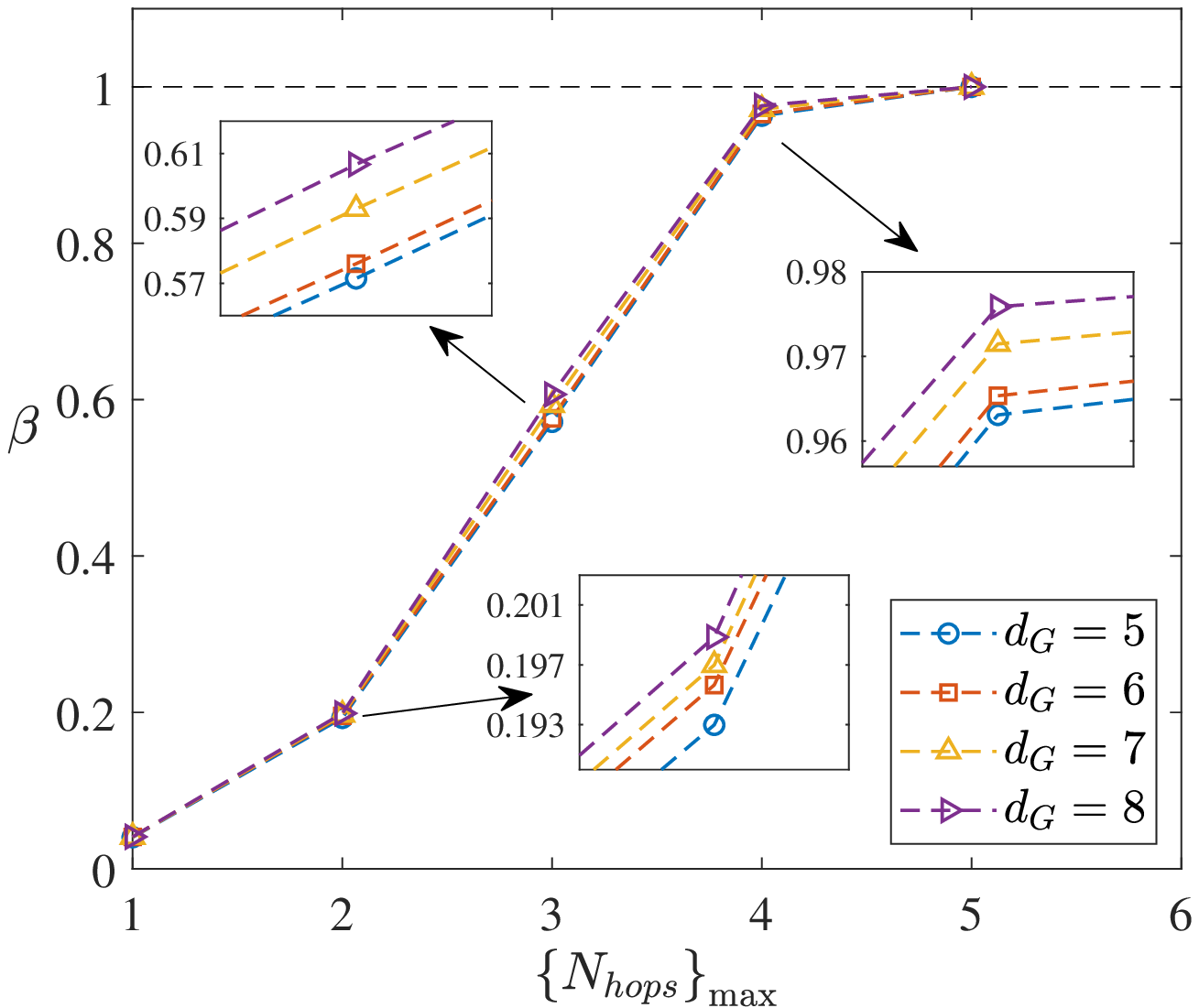}
		}
		\caption{Impact of $d_G$ for DWROSNs generated by PEIM algorithm: (a) Average node-to-node distance in each time slot, and (b) the relationship between node pair connectivity and maximum allowable relay hop count during time slot $\left[ {0,2000} \right)$.}
		\label{fig_5}
	\end{figure}

	We assume that $5\sim8$ LCTs are deployed on every GEO satellite, namely the node degree of GEO satellite nodes $d_G = 5\sim8$. 5 LCTs are deployed on every LEO satellite, namely the node degree of LEO satellite nodes $d_L = 5$. The DWROSNs are established based on the PEIM algorithm for each time slot under different $d_G$. The average node-to-node distance of the DWROSNs for different $d_G$ is shown in Fig. \ref{fig_5} (a). In each time slot, as the node degree increases, the average node-to-node distance of the DWROSNs decreases. During the entire operation time, the average value of $\overline H$ are 3.232, 3.218, 3.200, and 3.187 for node degree of GEO satellite nodes are 5, 6, 7, and 8, respectively.
	\begin{table*}[htb]
	\caption{node pair connectivity and wavelength demand of the DWROSNs generated by PEIM algorithm with different node degree during time slot $\left[ {0,2000} \right)$.\label{table_3}}
	\centering
	\renewcommand{\arraystretch}{1.2}
		\begin{tabular*}{500pt}{@{\extracolsep\fill}cllllllll@{\extracolsep\fill}}
			\toprule
			\multirow{2}{*}{\bm{${\left\{ {{N_{hops}}} \right\}_{\max }}$}} &
			\multicolumn{2}{c}{\bm{$d_G=5$}} &
			\multicolumn{2}{c}{\bm{$d_G=6$}} &
			\multicolumn{2}{c}{\bm{$d_G=7$}} &
			\multicolumn{2}{c}{\bm{$d_G=8$}}\\ \cline{2-9} 
			&
			\multicolumn{1}{c}{\bm{$\beta$}} &
			\multicolumn{1}{c}{\bm{$\overline {{N_\lambda }}$}} &
			\multicolumn{1}{c}{\bm{$\beta$}} &
			\multicolumn{1}{c}{\bm{$\overline {{N_\lambda }}$}} &
			\multicolumn{1}{c}{\bm{$\beta$}} &
			\multicolumn{1}{c}{\bm{$\overline {{N_\lambda }}$}} &
			\multicolumn{1}{c}{\bm{$\beta$}} &
			\multicolumn{1}{c}{\bm{$\overline {{N_\lambda }}$}} \\ \hline
			1 & 4.04\%  & 1.00   & 4.08\%  & 1.00   & 4.08\%  & 1.00  & 4.09\%  & 1.00  \\
			2 & 19.30\% & 11.75  & 19.57\% & 12.35  & 19.70\% & 13.10 & 19.89\% & 13.95 \\
			3 & 57.15\% & 54.55  & 57.60\% & 54.40  & 59.30\% & 57.50 & 60.67\% & 66.70 \\
			4 & 96.31\% & 124.40 & 96.53\% & 122.65 & 97.15\% & 118.00& 97.59\% & 112.60\\
			5 & 100.00\%& 132.75 & 100.00\%& 131.25 & 100.00\%& 124.75& 100.00\%& 118.05\\ \bottomrule
		\end{tabular*}
	\end{table*}

	Fig. \ref{fig_5} (b) shows the relationship between node pair connectivity and the maximum allowable relay hop count of the DWROSNs for different $d_G$ during time slot $\left[ {0,2000} \right)$. The DWROSNs with different $d_G$ reach full node pair connectivity within 5 hops. As the number of LCTs deployed on GEO satellite nodes increases, under the same maximum allowable relay hop count, the node pair connectivity of the DWROSNs increases. To simulate the wavelength demand of the DWROSNs, the traffic requests are generated and assigned to the DWROSNs following the RWA algorithm in Algorithm \ref{algorithm_2} ten times in each time slot. Table \ref{table_3} shows the details of the relationship between node pair connectivity and wavelength demand for the DWROSNs with different $d_G$ in time slot $\left[ {0,2000} \right)$. It can be clearly seen that the node pair connectivity and wavelength demand are a trade-off problem. Moreover, under the same $d_G$, the node pair connectivity and wavelength demand increase as the maximum allowable relay hop count increases. Under the same ${\left\{ {{N_{hops}}} \right\}_{\max }}$, as the $d_G$ increases, the node pair connectivity of the DWROSNs increases, and the wavelength demand of the DWROSNs decreases.
	\section{Conclusion}
	The inter-satellite links topology design problem of the DWROSN is studied in this paper. We have established a mathematical model for the dual-layer constellation, predicted the visibility of the ISLs for all node pairs, and the LAS based on the PEIM algorithm is proposed to solve the links assignment problem. We estimated the average node-to-node distance, node pair connectivity, wavelength demand, and transmission delay for full connectivity with all-optical wavelength routing in the dual-layer satellite network. The simulation results show that the DWROSN constructed by the PEIM algorithm has a short average node-to-node distance, and the PEIM algorithm reduces the average node-to-node distance by 7.6\% and 25.1\%, compared with the ACT and Greedy algorithms, respectively. The PEIM algorithm privets the long route path, and all node pairs could interconnect within 5 hops. Estimating traffic transmission performance, the DWROSN constructed by the PEIM algorithm shows a low wavelength demand and low average delay. The PEIM algorithm reduces the wavelength demand of the full connectivity DWROSN by 20.3\% and 64.9\%, compared with the ACT and Greedy algorithms, respectively. The simulation for the impact of the GEO layer node degree shows that the high $d_G$ improves the node pair connectivity and reduces the wavelength demand of the DWROSN constructed by the PEIM algorithm, and the node pair connectivity and the wavelength demand of the DWROSN is a trade-off problem. As can be seen from above, designing the topology of the ISLs with the PEIM algorithm in DWROSN is feasible, and the research in this paper means the reduction of the cost of on-board resources could be brought from not only route scheme optimization but also from designing the topology of the ISLs. From this perspective, it has an important reference for the design of the massive optical satellite network.
	\section*{Acknowledgment}
	This work is supported by the Free-space Optical Communication Technology Research Center.
	\subsection*{Conflict of interest}
	The authors declare no potential conflict of interests.
	%
	%
	\noindent
	\bibliography{mybibfile}%
\end{document}